\documentclass[12pt,preprint]{emulateapj}

\begin{document}

\title{Sw 1644+57/GRB 110328A: the physical origin and the composition of the relativistic outflow}

\author{\sc Lang Shao\altaffilmark{1,2,3}, Fu-Wen Zhang\altaffilmark{1,2,4}, Yi-Zhong Fan\altaffilmark{1,2} and Da-Ming Wei\altaffilmark{1,2}}
\altaffiltext{1}{Purple Mountain Observatory, Chinese Academy of Sciences, Nanjing 210008, China;yzfan@pmo.ac.cn,}
\altaffiltext{2}{Key Laboratory of Dark Matter and Space Astronomy, Chinese Academy of Sciences, Nanjing 210008, China,}
\altaffiltext{3}{Department of Physics, Hebei Normal University, Shijiazhuang 050016, China,}
\altaffiltext{4}{College of Science, Guilin University of Technology, Guilin, Guangxi 541004, China.}

\begin{abstract}
Sw 1644+57/GRB 110328A is a remarkable cosmological X-ray outburst detected by the {\it Swift} satellite. Its early-time ($t\lesssim 0.1$ days since the trigger) X-ray emission resembles some gamma-ray bursts (GRBs), e.g., GRB 090417B. But the late-time flaring X-ray plateau lasting $> 40$ days renders it unique. We examine the possibilities that the outburst is a super-long GRB powered either by the fallback accretion onto a nascent black hole or by a millisecond pulsar, and find out that these two scenarios can address some but not all of the main observational features. We then focus on the model of tidal disruption of a (giant) star by a massive black hole. The mass of the tidal-disrupted star is estimated to be $\gtrsim$ a few solar masses. A simple/straightforward argument for a magnetic origin of the relativistic outflow is presented.
\end{abstract}

\keywords{gamma ray burst: general---radiation mechanisms:
non-thermal---X-rays: general}

\setlength{\parindent}{.25in}

\section{INTRODUCTION}
Sw 1644+57/GRB 110328A was discovered on 2011 March 28 at 12:57:45 UT by the {\it Swift} satellite \citep{Burrows11}. It is coincident with an
optical source at redshift $z = 0.353$ \citep{Levan11}, as well as a radio source \citep{Zauderer11}. The total isotropic energy in X-ray and $\gamma-$ray is $\sim 10^{53}$ erg and the luminosity of the flaring X-ray afterglow is $\sim 10^{47}-10^{48}~{\rm erg~s^{-1}}$, both comparable to regular gamma-ray bursts (GRBs). However, the source was detectable and variable in Burst Alert Telescope (BAT) more than 40 hr after the initial trigger (BAT re-triggered for quite a few times), with peak brightness on the order of 200 mCrab \citep{Sakamoto11}. Moreover the flaring X-ray plateau holds up to $>40$ days. These features are rather peculiar and have not been reported in any GRBs before.

In this work, we first compare the X-ray emission of Sw 1644+57/GRB 110328A with some {\it Swift} GRBs and examine whether such an ultra-long outburst can be explained within the regular GRB framework.
We then focus on the model of tidal disruption of a star by a massive black hole. Within such a scenario, we propose a simple/straightforward argument for a magnetic origin of the relativistic outflow and estimate the mass of the tidal-disrupted star.

\section{Is Sw 1644+57/GRB 110328A a super-long GRB?}
\begin{figure}
 \begin{center}
 \includegraphics[width=0.6\textwidth]{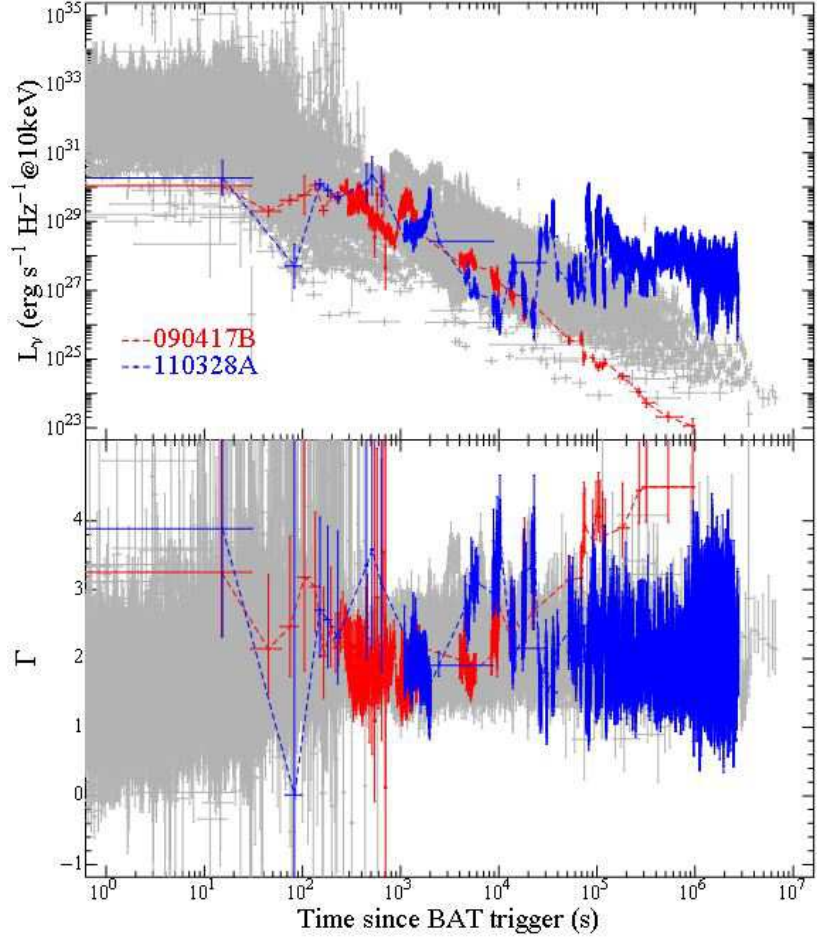}
 \end{center}
\caption{Evolution of the spectral luminosity (top) and the photon indices (bottom) at 10 keV of Sw 1644+57/GRB 110328A (blue) together with GRB 090417B (red) and other 137 long GRBs (gray) in their rest frames \citep{Evans10,Shao10}.}
\label{fig:X-ray}
\end{figure}

As a type of cataclysmic outbursts, GRBs have exhibited very diverse
characteristics in their observational properties. But for the X-ray (afterglow) emission, a canonical behavior has been established \citep[e.g.,][]{Zhang06}. Following \citet{Shao10}, in Figgure~\ref{fig:X-ray} we present the X-ray emission light curves of 138 long GRBs together with Sw 1644+57/GRB 110328A in their rest frames. Interestingly, for $t<0.1$ days since the trigger, Sw~1644+57/GRB 110328A {\it resembles} some specific GRBs, say, the super-long event GRB~090417B at a similar redshift ($z=0.345$). Moreover, both events exhibit spectral softening at $\sim10^4$~s and their optical emission is highly suppressed which may suggest very strong dust extinction \citep{Holland10,Bloom11}. Please see Table~\ref{tab:compar} for more details.

However, the late time temporal behavior of Sw~1644+57/GRB 110328A
makes it {\it unique}. For $t>10^{5}$ s the X-ray emission is strongly flaring and appears as a plateau. While for regular GRBs, the late time X-ray emission declines with time very quickly ($t^{-1.4}$ or so). As a result, at $t\sim 10^{6}$ s the X-ray flux of Sw~1644+57/GRB 110328A is brighter than that of any known GRBs by more than two orders of magnitude, in spite of the fact that the outburst is under-luminous when compared with other regular GRBs in the early phase (see Figure~\ref{fig:X-ray}).

The extremely-long duration of the flaring X-ray plateau imposes a stringent constraint on the physical origin \citep[see also][]{Dokuchaev11}. For long GRBs, the duration is usually governed by the activity of the central engine that is determined by the accretion process, depending on the size/structure of the progenitor star. As a very simple estimate, the prompt accretion has a duration comparable to the free fall timescale of the progenitor material $t_{\rm ff} \sim 40(1+z)R_{10}^{3/2}(M_{_{\rm BH}}/3M_\odot)^{-1/2}$ s. Please bear in mind that throughout this work, the convention $Q_{\rm x}=Q/10^{x}$ has been adopted except for some special notations. One needs a progenitor star with a size larger than $R \sim 10^{13}(M_{_{\rm BH}}/3M_\odot)^{1/3}$ cm and a number density profile roughly $dn/dR \propto R^{-3/2}$ to account for the ongoing X-ray plateau of Sw~1644+57/GRB 110328A \citep[e.g.,][]{Shao10}. Such a giant star however is {\it not} expected to launch an energetic relativistic outflow. Within the framework of GRB, the highly variable X-ray emission has been taken as the evidence of the prolonged-activity of the central engine \citep[e.g.,][]{Fan05,Zhang06,Wu07}, which could either be due to the fallback accretion onto the nascent black hole or alternatively the dipole radiation of a quickly rotating pulsar.

{\bf {\it In the fallback accretion scenario,}} the central engine can operate for a very long time. For example,
the long-lasting (up to $\sim 10^{6}$ s) but rather soft spectrum of the ``normally" declining X-ray afterglow of XRF 060218 calls for a central-engine-origin \citep[e.g.,][]{Fan06}. As found in the numerical simulation, the initial plateau given by the collapse of a GRB progenitor star only lasts $\leq 10^{3}$ s and the following fallback accretion rate can be approximated by $\dot{M} \propto t^{-5/3}$ \citep{MacFadyen01}. Such a temporal behavior is close to the X-ray afterglow decline of XRF 060218 but is significantly different from the flaring X-ray plateau of Sw 1644+57/GRB 110328A. In some specific cases, the emergence of the supernova reverse shock at the core at $\sim 10^{4}$ s can give rise to a prominent enhancement of the fallback-accretion rate, which lasts a few days and has a typical fallback accretion rate of $\sim {\rm a~few}\times 10^{-6}~{M_\odot~{\rm s^{-1}}}$ \citep{ZhangW08}. The energy output may be high up to $L \sim {\rm a~few}\times 10^{47}~{\rm erg~s^{-1}}~(\epsilon/10^{-3})(\theta_{\rm j}/0.1)^{-2}$, matching the X-ray luminosity of Sw 1644+57/GRB 110328A, where we assume an energy conversion coefficient $\epsilon \sim 0.001$ and a half-opening angle $\theta_{\rm j}\sim 0.1$ of the newly-launched outflow. However the second fallback accretion plateau is not long enough to account for the ongoing X-ray activity of Sw 1644+57/GRB 110328A.

{\bf {\it In the pulsar scenario,}} the central engine would not turn off until it has lost most of its rotation energy.  The relevant timescale can be quite long. Within the simplest dipole radiation model, the spin-down luminosity of the pulsar can be estimated by
\begin{eqnarray}
L_{\rm dip} = {4\times 10^{45}}~{\rm
erg~s^{-1}}~B_{\rm p,13}^2R_{\rm s,6}^6\Omega_{3.8}^4
\left(1+{(1+z)t\over \tau_{\rm 0}}\right)^{-2},
\label{eq:E_inj}
\end{eqnarray}
where $B_{\rm p}$ is the dipole magnetic field strength of the neutron
star at the magnetic pole, $R_{\rm s}$ is the radius of the pulsar, $\Omega$ is the angular frequency of rotation at $t=0$,
\[
\tau_{\rm 0}=4\times 10^{6}~{\rm s}~ (1+z)B_{\rm p,13}^{-2} \Omega_{3.8}^{-2}I_{45}
R_{\rm s,6}^{-6}
\]
is the corresponding spin-down timescale of the
pulsar, and $I\sim 10^{45}~{\rm g~cm^2}$ is the typical moment of
inertia of the pulsar \citep[e.g.,][]{Pacini67,Gunn69}. For $B_{\rm p,13}^{-2} \Omega_{3.8}^{-2}I_{45}
R_{\rm s,6}^{-6}>1$ (i.e., $\tau_{\rm 0}>5\times 10^{6}$ s), the long duration of Sw 1644+57/GRB 110328A
can be accounted for, while the spin-down luminosity given by Equation~(\ref{eq:E_inj}) is too low to be consistent with the observed X-ray luminosity $L_{\rm X}\sim 10^{47}~{\rm erg~s^{-1}}$. Unless, the outflow has been collimated into a narrow cone with a half-opening angle
\[
\theta_j
< 0.09~\epsilon_{{\rm x},-1}^{1/2}\Omega_{3.8}I_{45}^{1/2}L_{{\rm X},47}^{-1/2},
\]
where $\epsilon_{\rm x}$ is the radiative efficiency in the X-ray band and $L_{\rm X}$ is the observed luminosity of the long-lasting X-ray plateau. Yet, it is unclear whether such a narrow collimation can be achievable via the interaction with the expanding material of the associated-supernova. The other potential challenge of the pulsar model is how to produce the highly variable X-ray emission.

\section{Tidal disruption model}
Though the long-lasting activity of a GRB central engine seems possible, the luminous X-ray plateau lasting $>40$ days may favor a very different physical origin---``tidal disruption of a star by a massive black hole'' \citep[e.g.,][]{Hills75,Rees88,Ulmer99,LuY08}, which has already been applied to Sw 1644+57/GRB 110328A by \citet{de11} and \citet{Bloom11}.

\subsection{Tidal disruption model: a brief discussion}
The minimum variability timescale of the X-ray emission of Sw 1644+57/GRB 110328A is $\sim 78$ s \citep{Bloom11},
suggesting a black hole mass $M_{_{\rm BH}} \sim 7.8\times 10^{6}M_\odot/[3(1+z)]\sim 2\times 10^{6}M_\odot$, where $M_\odot$ is the solar mass \citep[e.g.,][]{LuY08}.
The tidal radius of a star captured by the massive black hole can be estimated by
\begin{equation}
R_{_{\rm T}} \sim 6.3\times 10^{13}~{\rm cm}~r_{*,1} m_{*,0.6}^{-1/3}M_{_{\rm BH,6.5}}^{1/3},
\end{equation}
where $m_*=M_*/M_\odot$, $r_*=R_*/R_\odot$ and $M_*$ ($R_*$) is the mass (radius) of the captured star.
Some material is unbound. The bound part may create flare as it accretes onto the black hole. The most bound material returns to the pericenter on a timescale \citep[e.g.,][]{Ulmer99}
\begin{equation}
t_{\rm fallback} \sim 4.4~{\rm days}~(1+z)(5R_{\rm p}/R_{_{\rm T}})^{3}r_{*,1}^{3/2}m_{*,0.6}^{-1}M_{_{\rm BH,6.5}}^{1/2},
\end{equation}
where $R_{\rm p}$ is the pericenter of the star's orbit\footnote{For a Schwarzschild (or a slowly rotating) black hole $R_{_{\rm T}}/R_{_{\rm S}}\sim 11.5r_{*}m_{*}^{-1/3}M_{_{\rm BH,6.5}}^{-2/3}$, the smallest distance to which the test particle on a parabolic orbit can approach and yet not be swallowed by the black hole is $R \sim 2R_{_{\rm S}}$, where $R_{_{\rm S}}$ is the Schwarzschild radius \citep{Kobayashi04}.}. Therefore the duration of the X-ray transient may be accounted for if the captured star is a (red) giant or alternatively an ``S-star'' as found in the Galactic center \citep[e.g.,][]{Alexander05}. The highly variable X-ray emission may reflect the instability involved in the accretion process.

\subsection{The relativistic movement: constraint on the physical process launching the outflow}\label{sec:magnetic}
The analysis of current observational data of Sw 1644+57/GRB 110328A suggests that the outflow is likely relativistic
with an initial bulk Lorentz factor $\Gamma_{\rm i} \gtrsim 10$ \citep{Bloom11}. If such an estimate can be confirmed by the late super-luminal expansion measurement, the outflow should be launched and then accelerated via some magnetic processes since the pure hydrodynamic acceleration is {\it disfavored}, as shown below.

A baryonic ejecta will be accelerated by the thermal
pressure until it becomes optically thin or saturates at a radius $R_{\rm
f}$, depending on the baryon-loading of the outflow. The
optical depth of the photon at the radius $R_{\rm f}$ can be estimated as
\citep[e.g.,][]{Pacz90,Jin10}:
\begin{equation}
\tau\sim\int^{\infty}_{R_{\rm f}}(1-\beta)n\sigma_{\rm T}dR\sim 1, \label{eq:M1}
\end{equation}
where $n\sim L/4\pi R^2\eta m_{\rm p}c^3$ is the number density of electrons coupled with
protons in the observer's frame, $\eta$ is the dimensionless entropy of the initial ejecta, $\sigma_{\rm T}$ is the Thompson cross section, $\beta$ is the velocity of the outflow in units of $c$ (the speed of light), and $m_{\rm
p}$ is the rest mass of protons. Combing with the relations $R_{\rm f}\sim\Gamma R_{0}$ \citep[e.g.,][]{Piran99} and
$\beta \sim1-1/2\Gamma^{2}$, Equation~(\ref{eq:M1}) gives
\begin{equation}
\frac{L\sigma_{\rm T}}{8\pi\eta\Gamma^{3}m_{\rm p}c^{3}R_{0}}\sim 1, \label{eq:M2}
\end{equation}
where $R_0\geq R_{_{\rm S}}$ is the initial radius of the outflow getting accelerated.
The final bulk Lorentz factor of accelerated-outflow is related to $L$ and $R_{0}$ as
\begin{equation}
\Gamma \leq  \Gamma_{_{\rm M}}=8.6 (\eta/\Gamma)^{-1/4}L_{49}^{1/4}R_{0,12}^{-1/4}\leq 8.6 L_{49}^{1/4}R_{0,12}^{-1/4},
\label{eq:Gamma_limit}
\end{equation}
where the fact that $\Gamma \leq \eta$ has been taken into account \citep[see also][]{Meszaros00}. Therefore the hydrodynamic process is unable to accelerate the outflow to a bulk Lorentz factor $\sim \Gamma_{\rm i}\geq 10$, which in turn suggests that the outflow is not baryonic. Actually it is unlikely to launch an energetic baryonic outflow with $\eta \gtrsim {\rm a~few}$ for an accreting massive black hole ($M_{_{\rm BH}} \sim 10^{6-7}M_\odot$). The reason is that the inner region of the surrounding disk is very cool with a temperature
$T_{\rm in} \sim 1~{\rm keV}~\dot{M}_{-6}^{1/4}h_{-1}^{-1/4}R_{12}^{-1/2}$, where $h$ is the ratio between the height and the radius of the disk material. The cooling rate per unit volume owing to neutrinos ($\propto T_{\rm in}^{11}$) in such a ``cold" region can be ignored. Hence no significant neutrino emission is expected. Any other outflows driven by the thermal pressure from the disk suffer from significant baryon loading and cannot have an $\eta$ significantly larger than 1.

Consequently we conclude that the relativistic outflow from a black hole with a mass $\sim 10^{6-7}~M_\odot$ cannot be baryonic. The magnetohydrodynamic (MHD) processes are needed to launch and then accelerate the relativistic outflow. Our argument is in agreement with the previous conclusion based on numerical simulations \citep[e.g.,][]{Meier01,Vlahakis04}. The flaring X-ray/infrared emission may be highly polarized if these photons were the synchrotron radiation of electrons accelerated by the magnetic reconnection in the relativistic outflow. \citet{Bower11} set a 2-sigma upper limit on the linear polarization fraction of $4.5\%$ at a frequency of 8.4 GHz.
Such a low linear-polarization degree favors an external-shock origin of the radio emission, as suggested by \citet{Bloom11}. This is because in the external-shock model the electrons are accelerated at the forward shock front and the magnetic field is shock-generated/random. The linear polarization of the synchrotron radiation of the electrons cancels with each other and the net linear polarization degree is expected to be very low.

\subsection{Estimating the mass of the tidal-disrupted star}
Till 2011 May 12, the isotropic-equivalent X-ray/$\gamma-$ray emission energy is $E_{\rm iso}\gtrsim 10^{53}~{\rm erg}$. Such a huge amount of energy sheds some light on the mass accreted onto the central black hole. The possible physical origins of the X-ray emission have been discussed in some detail by \citet{Bloom11}. Instead of going into detail of these possibilities, here we make the {\it simplest} assumption and then estimate the mass of the captured star.

{\it Case-I: the X-ray emission is mainly from the disk}. In order of magnitude, the total disk luminosity
is \citep[e.g.,][]{Shapiro83}
\begin{equation}
L \lesssim 0.1 \dot{M}c^{2} \sim 2\times 10^{47}~{\rm erg~s^{-1}}~\dot{M}_{-6},
\end{equation}
where $\dot{M}$ is the accretion rate (in units of $1~M_\odot~{\rm s^{-1}}$).
Please note that the disk emission is not narrowly-collimated and just about half of the star's material is bound, to account for the ongoing X-ray emission one needs a tidal-disrupted star with a mass
\[
M_* \gtrsim E_{\rm iso}/0.1~c^{2}  \gtrsim {\rm a~few}~M_\odot.
\]

{\it Case-II: the X-ray emission is mainly from the relativistic outflow}. As shown in \S\ref{sec:magnetic}, the relativistic outflow should be launched by some MHD processes. In the most widely discussed Blandford-Znajek process \citep{BZ77,Lee01}, the luminosity of the electromagnetic outflow can be estimated by $L_{_{\rm BZ}}\sim 1.8\times 10^{45}~{\rm erg}~(a/0.4)^{2}M_{_{\rm BH,6.5}}^{2}B_{\rm H,7}^{2}$, where $a$ is the spin parameter of the massive black hole\footnote{This parameter is hard to estimate/constrain. Recently, \citet{Kato10} found that the spin parameters of black holes in Sgr A* and in Galactic X-ray sources have a unique value of $a\sim ~0.44$. Hence we normalize the poorly-known $a$ to a value $\sim 0.4$.},
$B_{_{\rm H}}\sim 1.1\times 10^{7}~{\rm G}~\dot{M}_{-6}^{1/2}R_{_{\rm H},12}^{-1}$ is the magnetic field strength on the horizon and $R_{_{\rm H}}=(1+\sqrt{1-a^{2}})R_{_{\rm S}}/2$. With a proper collimation, the X-ray emission could have a luminosity
\begin{equation}
L_{\rm X} \sim 2\epsilon_{\rm x}L_{_{\rm BZ}}/\theta_{\rm j}^{2} \sim 10^{47}~{\rm erg~s^{-1}}~\epsilon_{\rm x,-1}(a/0.4)^{2}\dot{M}_{-5.5}\theta_{\rm j,-1}^{-2}.
\end{equation}
Again, one needs $M_* \geq {\rm a~few}~M_\odot$ to account for the data of Sw 1644+57/GRB 110328A.

\section{Conclusion and Discussion}
Sw 1644+57/GRB 110328A was detectable and variable in BAT more than 40 hr after
the initial trigger (the BAT re-triggered for quite a few times). The early time ($t\lesssim 0.1$ day since the trigger) X-ray emission of this cosmological outburst seems not unusual, which actually mimics GRB 090417B. But the late-time flaring X-ray plateau lasting $>40$ days renders it remarkable and disfavors the speculation that it is a super-long GRB powered either by the fallback accretion onto the nascent black hole or by a millisecond pulsar.
A plausible model is the tidal disruption of a (giant) star by a massive black hole, as widely speculated since 1970s. In such a scenario with a simple/straightforward argument we show that the relativistic outflow should be launched by some MHD processes and the mass of the tidal-disrupted star is estimated to be $\gtrsim$ a few solar masses.

We note that the tidal-disrupted object could also be a binary instead of a single star, given that the high stellar density prevails at the center of a galaxy \citep[e.g.,][]{Alexander05} and about half of the stars form in binaries \citep[e.g.,][]{Duquennoy91}. Significant variability is naturally expected since the stars with different radii may have very different $t_{\rm fallback}$ and then give rise to separate flares. For the current data rich of substructures this possibility may not be ruled out. However the rate of the tidal disruption of both stars in the binary might be about 10 times smaller than the rate of the tidal disruption of only one star, with the other ejected with a high velocity \citep[e.g.,][]{Ginsburg06,Ginsburg07}. Even so, the latter would probably suffer strong tidal perturbations and mass loss before ejection \citep{Antonini10}.

Though Sw 1644+57/GRB 110328A is likely powered by an accreting massive black hole, the early ($t<10^{4}$ s) X-ray emission, including the luminosity, the spectral evolution, and the temporal behavior, is rather similar to that of GRB 090417B, for which the central engine is plausibly a stellar black hole. This fact suggests that {\it very different central engines can produce rather similar X-ray outbursts in a selected time interval,} possibly {\it via the same kind of energy extraction process}.

In this work we did not investigate the forward shock emission of the relativistic outflow. Since the central engine does not turn off yet, any reliable calculation should take into account the energy injection of the outflow as well as the cooling of the forward shock electrons by the photons generated in the internal energy dissipation (the byproduct is a high energy emission component due to the inverse Compton scattering). A self-consistent approach can be found in \citet{Fan08}.

\section*{Acknowledgments}
We thank the anonymous referee for helpful comments. This work made use of data supplied by the UK {\it Swift} Science Data Centre at the University of Leicester. This work was supported in part by a special grant from Purple Mountain Observatory, the National Natural Science Foundation of China (grants 10973041, 10621303, and 11073057) and Chinese Academy of Sciences, and National Basic Research Program of China (grants 2007CB815404 and 2009CB824800). F.-W.Z. acknowledges the support by Guangxi Natural Science Foundation
(2010GXNSFB013050) and the doctoral research foundation of Guilin University of Technology.

\clearpage

\begin{table*}
 \begin{minipage}{140mm}
 \centering
  \caption{Comparison Between Sw 1644+57/GRB 110328A and GRB 090417B. }
\label{tab:compar}
  \begin{tabular}{llllllll}
  \hline

GRB & z  & Duration           &  Photon index  & $N_{\rm H}$ (Host Galaxy)           & $A_{v}$ &  Ref.\\
    &    & (s) (15-150 keV)     &   (15-150 keV) & $ {\rm cm^{-2}} $   & mag     & \\
\hline
 090417B  &0.345 & $>2130$    & 1.89$\pm$0.12 &$(1.1-2.4)\times10^{22}$ & $\gtrsim 12$  &1 \\
 110328A  &0.353 & $>10^{5}$   & 1.72$\pm$0.18  &$\sim 2\times10^{22}$    & $\gtrsim 4.5-10$  & 2,3,4 \\

\hline

\end{tabular}

References. (1) Holland et al. 2010; (2) Bloom et al. 2011; (3) Burrows et al. 2011; (4) Levan et al. 2011.
\end{minipage}

\end{table*}

\end{document}